# Thin-Film Lithium Niobate Acoustic Filter at 23.5 GHz with 2.38 dB IL and 18.2% FBW

Omar Barrera, Sinwoo Cho, Lezli Matto, Jack Kramer, Kenny Huynh,
Vakhtang Chulukhadze, Yen-Wei Chang, Mark S. Goorsky, and Ruochen Lu

*Abstract*—This work reports an acoustic filter at 23.5 GHz with a low insertion loss (IL) of 2.38 dB and a 3-dB fractional bandwidth (FBW) of 18.2%, significantly surpassing the state-of-the-art. The device leverages electrically coupled acoustic resonators in 100 nm 128° Y-cut lithium niobate (LiNbO$_3$) piezoelectric thin film, operating in the first-order antisymmetric (A1) mode. A new film stack, namely transferred thin-film LiNbO$_3$ on silicon (Si) substrate with an intermediate amorphous silicon (a-Si) layer, facilitates the record-breaking performance at millimeter-wave (mmWave). The filter features a compact footprint of 0.56 mm$^2$. In this letter, acoustic and EM consideration, along with material characterization with X-ray diffraction and verified with cross-sectional electron microscopy are reported. Upon further development, the reported filter platform can enable various front-end signal-processing functions at mmWave.

## I. Introduction

COMPACT radio frequency (RF) front-end filter components are indispensable in mobile devices. Among various filter technologies [1], acoustic resonator-based filters are unique, providing four orders of magnitude smaller sizes compared to electromagnetic (EM) alternatives along with better frequency selectivity [2]–[4]. Acoustic resonators convert electrical signals to mechanical vibrations and vice versa through piezoelectricity [5]. Thus, one can process electrical signals in acoustics, while benefiting from shorter wavelengths and better energy confinement [4]. In incumbent RF front ends, thin-film bulk acoustic wave resonators (FBAR) and surface acoustic wave (SAW) devices are the dominant filtering solutions [6], [7]. A significant amount of commercial and academic success has been demonstrated in the sub-6 GHz frequency band using the aforementioned acoustic devices using aluminum nitride/scandium aluminum nitride (AlN/ScAlN) [8]–[10] and lithium niobate/lithium tantalate (LiNbO$_3$/LiTaO$_3$) [11]–[15]. With the ever-growing demand for wireless bandwidth, RF front-end devices keep developing into millimeter-wave (mmWave) bands, toward higher frequency and wider fractional bandwidth (FBW). Hence, scaling acoustics into mmWave will play a monumental role in

Manuscript received 1 June 2023; revised XX June 2023; accepted XX June 2023. This work was supported by DARPA COmpact Front-end Filters at the ElEment-level (COFFEE).

O. Barrera, S. Cho, J. Kramer, V. Chulukhadze, Y.-W. Chang, and R. Lu are with The University of Texas at Austin, Austin, TX, USA (email: omarb@utexas.edu). L. Matto, K. Huynh, and M. Goorsky are with University of California Los Angeles, Los Angeles, CA, USA.

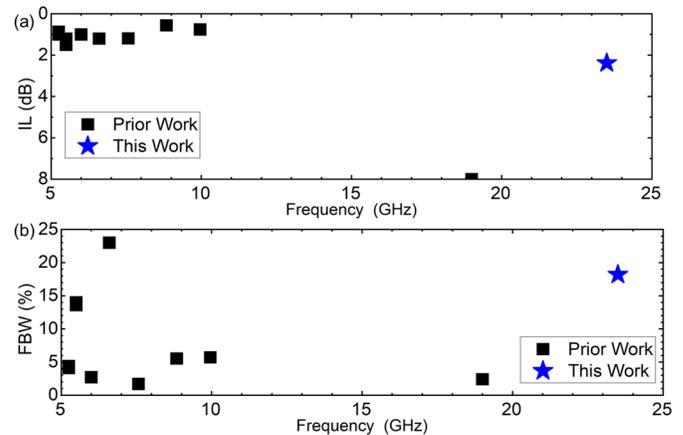

Fig. 1 Survey of (a) IL and (b) FBW in acoustic filters above 5 GHz.

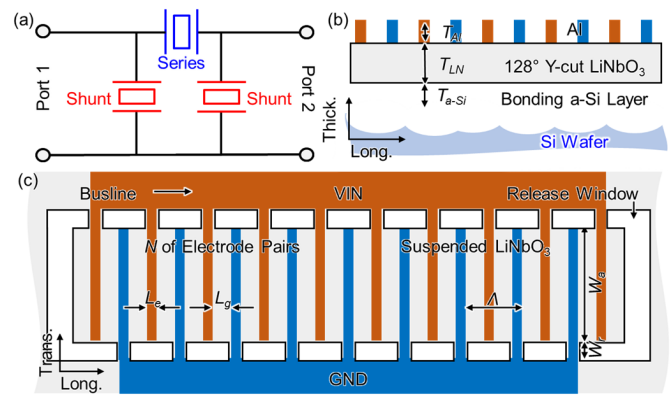

Fig. 2 (a) Filter circuit, and resonator (b) top and (c) side schematics.

providing low-loss and compact filtering solutions [16].

Nevertheless, achieving mmWave acoustic filters is not trivial [survey in Fig. 1 (a)(b)], due to the limited availability of high-performance acoustic resonators at mmWave. One frequency scaling method uses smaller feature sizes or thinner films for wavelength reduction [17]–[21], which inevitably suffers from excessive loss, i.e., lower quality factor (*Q*), and higher insertion loss (IL), due to its fabrication difficulties and attenuation in ultra-thin film. An alternative frequency scaling method uses higher-order modes [22], [23] However, this method is intrinsically limited by reduced electromechanical coupling ($k^2$), resulting in higher IL and lower FBW.

More recently, first-order antisymmetric (A1) mode resonators have been demonstrated at mmWave, using thin-film LiNbO$_3$ transferred on low-loss substrates with intermediate

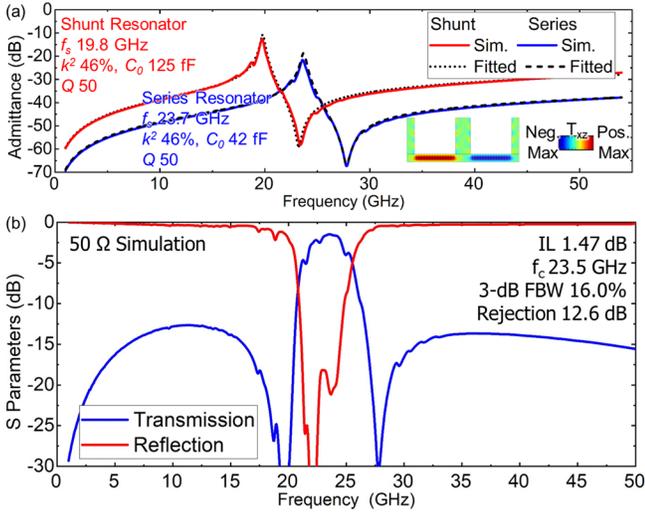

Fig. 3 (a) Simulated admittance amplitude, inset key specifications and vibration mode shape. (b) Simulated filter response.

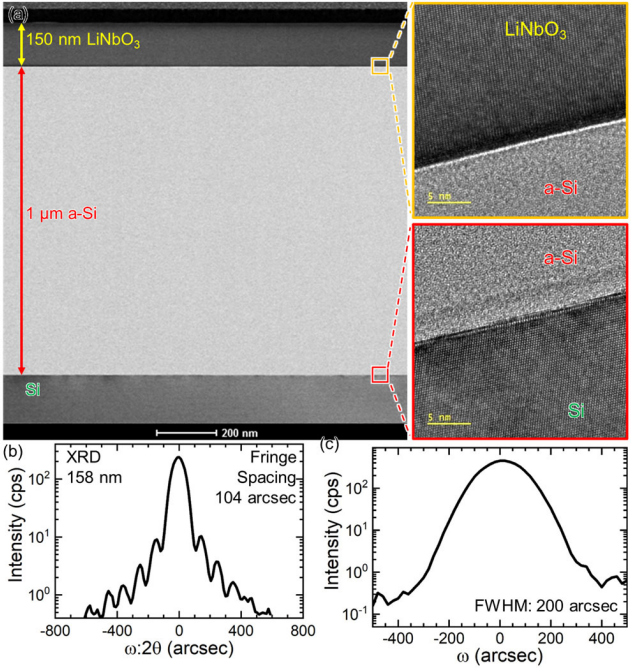

Fig. 4 (a) Cross-sectional TEM image of the stack and interfaces. (b) XRD symmetric ω:2θ scan of the (01$\bar{1}$4) reflection. (c) XRD symmetric rocking curve of the (01$\bar{1}$4) reflection.

amorphous silicon (a-Si) [24], [25]. This resonator platform proposed in this work, combined with the EM-acoustic co-design, is demonstrated to be the key to unlocking mmWave acoustic filters. In this work, we present a third-order filter at 23.5GHz in thin-film LiNbO$_3$. The design exhibits an IL of 2.38dB and an FBW of 18.2%, surpassing the state-of-the-art (SoA) (Fig. 1). The filter has a small footprint of 0.56 mm². The significant improvement demonstrated in these results paves the way towards next-generation mmWave acoustic filters.

## II. DESIGN AND SIMULATION

The filter topology comprises one series resonator and two identical shunt resonators [Fig. 2 (a)]. The resonator structure contains an array of interdigital transducers (IDT) on top of a 128° Y-cut LiNbO$_3$ thin-film, suspended over a silicon (Si) substrate with 1 µm thick a-Si bonding layer [Fig. 2 (b)]. Si is selected as the sacrificial layer for its ease of mechanical suspension. The electric field generated between IDT fingers excites the A1 mode through the $e_{15}$ piezoelectric coefficient. The resonator design follows that in [24], using aluminum for electrodes. Additionally, a thick metal layer acts as a natural heatsink and lowers electrical loss. A list of key design parameters is labeled in Fig. 2 (c) and listed in Fig. 5.

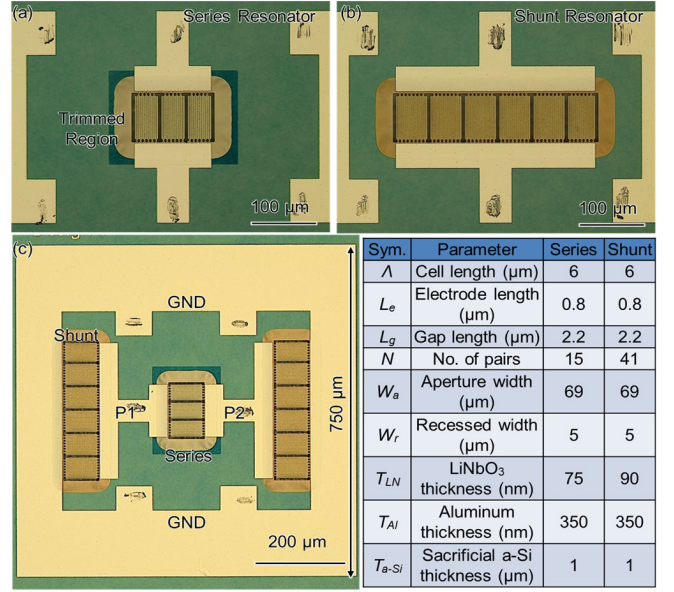

Fig. 5 Microscopic images of fabricated (a) series resonator, (b) shunt resonator, and (c) filter. Key dimensions are listed in the table.

COMSOL finite element analysis (FEA) simulated A1 resonator admittance is plotted in Fig. 3 (a), showing high $k^2$ of 46%. The series and shunt resonators are designed with different film thicknesses (more specifically, 75 nm for series and 90 nm for shunt) to achieve the necessary shift in resonance ($f_s$) for the 23.5 GHz bandpass filter [26]. The static capacitance $C_0$ is chosen to match the impedance of the filter to 50 Ω while providing 12 dB out-of-band (OoB) rejection. The resonator $Q$ is assumed to be 50 from earlier results [27]. The simulated resonator admittance is imported into Keysight Advance Design System for filter synthesis. The simulated filter [Fig. 3 (b)] shows an IL of 1.47 dB, a center frequency ($f_c$) of 23.5 GHz, a 3-dB FBW of 16%, and an OoB of 12.6 dB. The low-loss and wideband filter response validates the promising perspective of the thin-film LiNbO$_3$ platform at mmWave.

## III. FABRICATION AND MEASUREMENT

The stack for this work (LiNbO$_3$-aSi-Si) is provided by NGK Insulators Ltd. The thickness of the stack is first validated using transmission electron microscopy (TEM) imaging [Fig. 4 (a)]. The thickness of the LiNbO$_3$ layer is around 150nm, and the bonding a-Si layer is 1µm. Good interfaces between LiNbO$_3$, a-Si, and Si are observed. Next, more quantitative material analysis is carried out using X-ray diffraction (XRD). A plot of the triple-axis diffractometry (TAD) [29], [30] symmetric ω:2θ of (01$\bar{1}$4) and rocking curve of the piezoelectric layer are displayed in Fig 4 (b) and (c), respectively. The XRD measured a thickness of 158nm of the LiNbO$_3$ layer, revealing slight thickness variation across the sample. A full width at half maximum (FWHM) of 200 arcsec from the rocking curve is slightly worse than that with the sapphire substrate (60 arcsec in [24]), but significantly improved from that without a-Si [28], corroborating the stack choice.

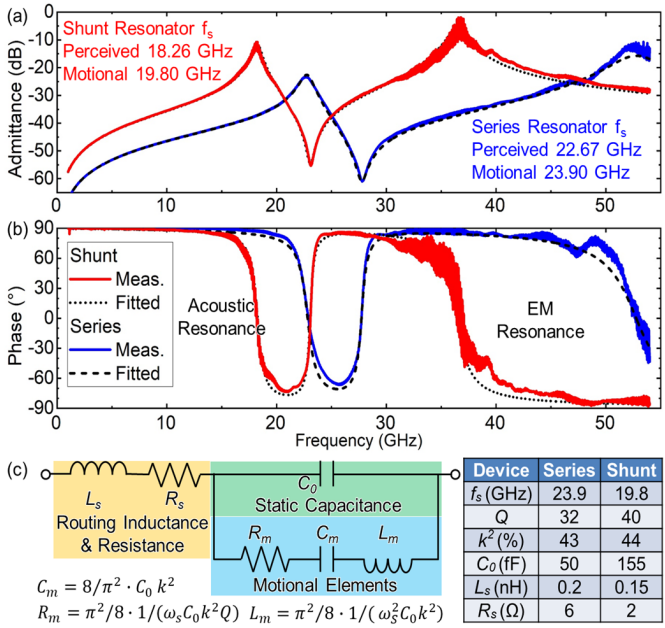

Fig. 6 Measured wideband admittance response in (a) amplitude and (b) phase. (c) Modified mmWave MBVD model and extracted key resonator specifications.

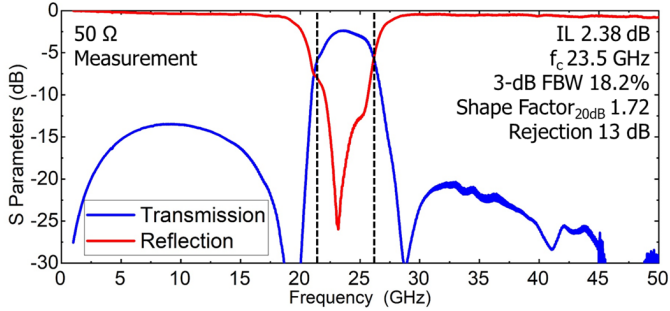

Fig. 7 Measured filter wideband transmission and reflection.

The fabrication process starts by trimming down the LiNbO$_3$ layer to 90 nm (the thickness of the shunt resonator) over the entire surface of a 2.1 by 1.9 cm sample. This is accomplished using ion beam-assisted argon gas cluster trimming, which is reported to maintain surface roughness and high crystallinity [29]. The trim process using ion-beam has been precisely characterized in-house, validated by atomic force microscope (AFM). Next, release windows are defined and etched using the ion beam. Since the resonator bank dimensions are large, resonators are divided into multiple subsections for easier release. Moreover, small release windows at the end of the IDT's help confine acoustic energy and slightly improve performance. Afterward, another trimming to 75nm is performed over selected regions for series resonators, providing the required $f_s$ shift. Finally, metal electrodes are patterned, and the resonators are released through xeon difluoride (XeF$_2$) Si etch. Optical images of the fabricated stand-alone series and shunt resonators and the filter are displayed in Fig. 5 (a), (b), and (c). The key dimensions are listed in Fig 5, the number of IDT pairs refer to the entire resonator. The filter has a small footprint of 0.75 mm by 0.74 mm, including the ground traces.

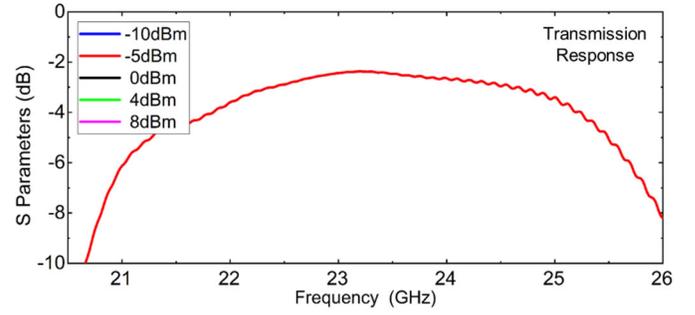

Fig. 8 In-band power handling response of the filter from −10 to 8 dBm.

The resonators and the filter are first measured using a Keysight vector network analyzer (VNA) in air at −15 dBm power level. The admittance and phase of the resonators are plotted in Fig. 6 (a) and (b), fitted with the mmWave MBVD circuit model [24] in Fig 6 (c). Unlike conventional MBVD models, the inductive effects from routing inductance $L_s$ are indispensable. More specifically, a higher frequency resonance of electromagnetic (EM) nature occurs due to the self-resonance of the reactive parasitics embedded in the resonator routing. $L_s$ and routing resistance $R_s$ are fitted based on EM resonances. Another effect is that the perceived resonances are now at 18.26 GHz for the series and 22.67 GHz for the shunt, which deviate from the mechanical resonances represented by the motional elements $L_m$, $C_m$, and $R_m$. The measurements show $Q$ around 40 and a high $k^2$ of around 42% (Fig. 6 inset table). $Q$ is defined at the anti-resonance due to the inclusion of $R_s$ and $L_s$.

The measured filter response is shown in Fig. 7. The 23.5 GHz filter exhibits a low IL of 2.38 dB IL, a wide 3-dB FBW of 18.2%, a 20 dB Shape Factor of 1.72 and an OoB rejection of 13 dB, matching device simulation. Compared with SoA low-loss acoustic filters (Fig. 1), this work shows significant frequency scaling and FBW enhancement. The reduced OoB performance is a drawback of using a low order filter with only 3 resonators, and will be improved in future works.

Finally, filter response at different power levels is characterized. The device is measured using the same VNA at several power level intervals between −10 to 8 dBm (maximum available power), showing little difference in the transmission (Fig. 8). We suspect thermal nonlinearity will be the dominant factor for such thin-film devices if larger input powers are applied [30] . Further study on linearity and power handling will be studied in future works.

## IV. Conclusion

We report the design and implementation of a 23.5 GHz acoustic filter with a low IL of 2.38 dB and a 3-dB FBW of 18.2%, using 100 nm LiNbO$_3$ on Si with an intermediate amorphous Si layer. The device has been tested under different power conditions showing good linear behavior.

The filter performance significantly surpasses SoA and highlights the possibility of scaling compact acoustic front-end signal processing toward mmWave bands.